\pdfoutput=1
\documentclass{JINST}
\usepackage{graphicx}
\usepackage{caption}
\usepackage{subcaption}
\usepackage{amsmath}
\usepackage{mathtools}
\usepackage{hepunits}
\newcommand{\bigcell}[2]{\begin{tabular}{@{}#1@{}}#2\end{tabular}}

\title{Optimizing the Performance of a High-Granularity Silicon-Pad EM Calorimeter}
\author{Stathes Paganis$^c$\thanks{Corresponding
author.}~, Andreas Psallidas$^c$, Arnaud Steen$^c$\\
\llap{$^c$}Department of Physics, National Taiwan University,\\
No 1, Sec 4, Roosevelt Road, Taipei 10617, Taiwan\\
  E-mail: \email{stathes.paganis@cern.ch}}

\abstract{
   A silicon-based fine granularity calorimeter is a
   potential technology for the future International Linear Collider ILC, 
   the future circular collider CEPC, and is also the chosen technology for the upgraded CMS experiment of the Large Hadron Collider. Active silicon sensing pads are used as MIP counters and the
   standard calibration of the calorimeter uses weights based on the average energy 
loss, $dEdx$. In this work, the limitations of the
   $dEdx$ calibration method in terms of energy linearity, scale and resolution are
   explored. In the case of a calorimeter with varying passive layer thickness as the one planned for CMS, the $dEdx$ method leads to a significant constant term in the resolution function and a non-linearity of energy response. For these reasons, a method based on the calorimeter sampling fraction that exploits the per-event measured shower depth is presented and 
shown to deliver superior absolute energy scale, linearity and resolution. Calorimetric designs in which the back of the shower is sampled less, offer reduced cost without loss in performance. Therefore, a proper calibration as proposed here is crucial in obtaining the most cost- and performance-effective silicon-sampling calorimeter design.
}

\keywords{Calorimeters, Calorimeter Methods}

\begin{document}
\bibliographystyle{plain}  
\section{Introduction}

The next step in electromagnetic (EM) and hadronic calorimetry of high
energy showers is a device that can provide fine 3D shower development information
for a reasonably good energy resolution and linearity. A popular design is a sampling
calorimeter of 0.5 to 1~cm$^{2}$ area silicon pads as the active layers and tungsten plates as 
passive layers. A silicon-strip tungsten electromagnetic calorimeter of this type 
has already been constructed and commissioned for the PAMELA satellite mission 
experiment \cite{pamela}. However, a silicon-pad sampling calorimeter with fine granularity 
and timing capability has not been built yet.
Such a device has been proposed by the CALICE
collaboration of the International Linear Collider (ILC), \cite{Calice} \cite{ilc},
for which prototypes have been built and tested in test beams. The same
technology has been proposed for the Circular EP Collider (CEPC) \cite{cepc}, 
and recently for the phase-2 endcap calorimeters of CMS \cite{cms}. The CMS
collaboration at the LHC has already approved the HGCal detector and
intense R\&D has commenced. The design of HGCal has not yet been finalized 
and an issue is always the reduction of cost for a minimal loss of performance.
Various factors such as space, cooling, powering of the detector, 
also provide constraints to the final design. Similar restrictions will exist in 
future experiments, so the present work explores the potential problems 
and compromises in EM energy reconstruction that alternative designs 
can lead to. Our aim is to address the impact of alternative and cost-effective 
designs on the energy resolution stochastic and constant terms.
As an example, if a design with fewer sampling layers in the 
back of the calorimeter leads to practically the same performance with a uniform/homogeneous sampling
calorimeter, then the cost reduction may be significant. Therefore, 
design optimization should not be based on performance measures obtained from
methods that introduce biases. Exploiting the information provided by a 3D 
sampling calorimeter, such as shower shapes, may allow to recover any potential 
compromises in performance in the case of more cost-effective designs.\\

\noindent
In this work, besides studying the standard calibration method
(the so-called $dEdx$ method), 
we also explore an alternative calibration method based on 
the average calorimetric sampling fraction ($SF$) \cite{Wigmans}, and
compare its performance with the $dEdx$ method.
We first study the performance of a homogeneous silicon 
calorimeter (a calorimeter where the passive layers are identical) 
using the two methods. Subsequently, 
we examine a more complex design that is a candidate for the CMS experiment 
HGCal detector planned for LHC phase 2. The design is using a 
varying passive layer thickness, typically increasing with depth. 
We will call this design {\it inhomogeneous} and will show that the $dEdx$ method
has significant performance issues, while the $SF$ method retains its excellent performance
seen in the homogeneous Si-W calorimeter case. A concern about the simple $dEdx$ method, is that even in the ideal case of a noiseless detector, it generates a residual constant term of about 0.5\%. Such a large constant term is a concern for physics analysis, since at the forward region in LHC, the energies expected are at the few hundred \GeV~level where the constant term is dominant. An additional concern is that the $dEdx$ method leads to large EM energy non-linearity. In this work, a simple event-by-event calibration procedure that can mitigate these problems is introduced.

\section{Detector Simulation}

The detector used in this study is a 30-layer sampling calorimeter with passive layers 
made of layers of W, Pb and copper and active layers made of 300~$\mu$m thick 
silicon (Si) positioned on a PCB. This realistic design is simulated using the Geant4 
package \cite{Geant4}. We examine two different calorimetric designs:
\begin{itemize}
\item
a CALICE-like design of a homogeneous 
calorimeter with identical passive-active layers consisting of tungsten (W), silicon (Si), PCB and air. 
A W passive layer is of 0.86 radiation lengths ($X_0$), while the PCB is 1 mm thick and the air gap between layers is 2.5~mm.
\item
An inhomogeneous (non-uniform) calorimeter inspired by the LHC phase-2 CMS experiment proposal~\cite{Magnan}, presented in Table~\ref{t1}.
\end{itemize}
Both calorimeters consist of 30 20$\times$20~cm ($XY$) layers and have the same total radiation 
length.
In this work, the EM showering is modelled by the 
QGSP\_FTFP\_BERT physics list. Systematic effects 
due to incomplete description of the EM showers by the simulation 
can be studied with data. A validation of a number of 
physics lists including QGSP\_FTFP\_BERT  was performed 
in a 2016 test beam of a CMS HGCal 27$X_0$-deep prototype 
for a wide range of electron beam energies \cite{Jain}. The prototype 
sampling layer passive/active material composition 
is similar to the one used here. According to these 
studies, the simulation, when compared with data exhibits a linear 
behaviour, while the predicted absolute energy scales vary.\\

\begin{table}[hbt]
\caption{Layout of a realistic inhomogeneous 30-layer silicon electromagnetic calorimeter, inspired by similar designs proposed by the CMS collaboration \cite{Magnan}}
\label{t1}
\begin{tabular}{|l|l|}
\hline
\multicolumn{2}{|c|}{Inhomogeneous calorimeter} \\
\hline
Thermal shielding & 2mm Aluminium+26mm Foam+2mm Aluminium \\
\hline
Layer 1 & 0.5mm Cu+2mm Air+1.2mm FR4+0.3mm Si+3mm Cu+1mm Pb \\
\hline
5 modules & \bigcell{l}{1.75mm W+0.5mm Cu+2mm Air+1.2mm FR4+0.3mm Si \\ 3mm Cu+1mm Pb+3mm Cu+0.3mm Si+1.2 FR4+2mm Air+0.5mm Cu} \\
\hline
5 modules & \bigcell{l}{2.8mm W+0.5mm Cu+2mm Air+1.2mm FR4+0.3mm Si  \\ 3mm Cu+2.1mm Pb+3mm Cu+0.3mm Si+1.2 FR4+2mm Air+0.5mm Cu} \\
\hline
4 modules & \bigcell{l}{4.2mm W+0.5mm Cu+2mm Air+1.2mm FR4+0.3mm Si  \\ 3mm Cu+4.4mm Pb+3mm Cu+0.3mm Si+1.2 FR4+2mm Air+0.5mm Cu} \\
\hline
Layer 30 & 4.2mm W + Layer 1 \\
\hline
\end{tabular}
\end{table}

\noindent
Our goal is to isolate and study the impact of calibration on the ultimate 
resolution, linearity and energy scale, therefore,
the studies presented here involve a range of electron beams of constant energy 
fired from a single point along the beam axis $Z$ at the center of the detector. 
The analysis is performed at the hit level without 
any noise introduced to the measured energies in the active elements. 
During the simulation we record the energy deposited in passive layers, 
the energy leaking from the side, and the energy leaking from the back.
The leaking energy information per event can be added back to the reconstructed 
energy, thus allowing to study its impact on resolution, linearity and energy scale.
In this work, the leaking energy is assumed to be known exactly. 
Therefore, the leakage correction in the text corresponds to adding 
the actual leaking energy back to the reconstructed energy.

\begin{figure}[htb]
\begin{center}
\resizebox{0.70\textwidth}{!}{
\includegraphics{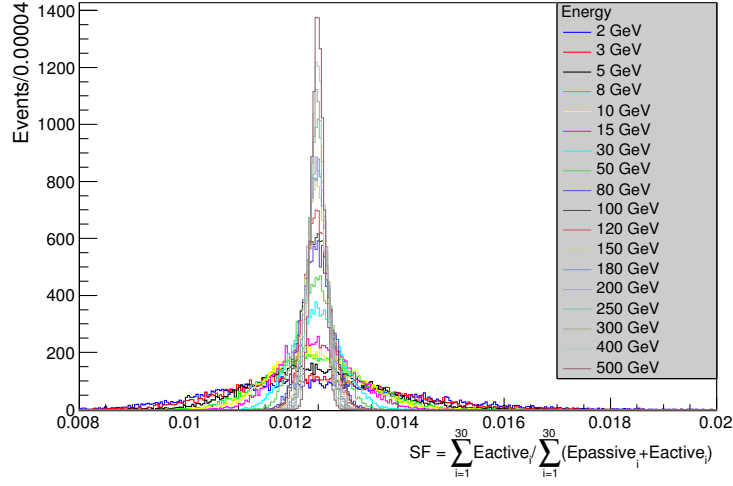}
}
\end{center}
\begin{center}
\resizebox{0.72\textwidth}{!}{%
\includegraphics{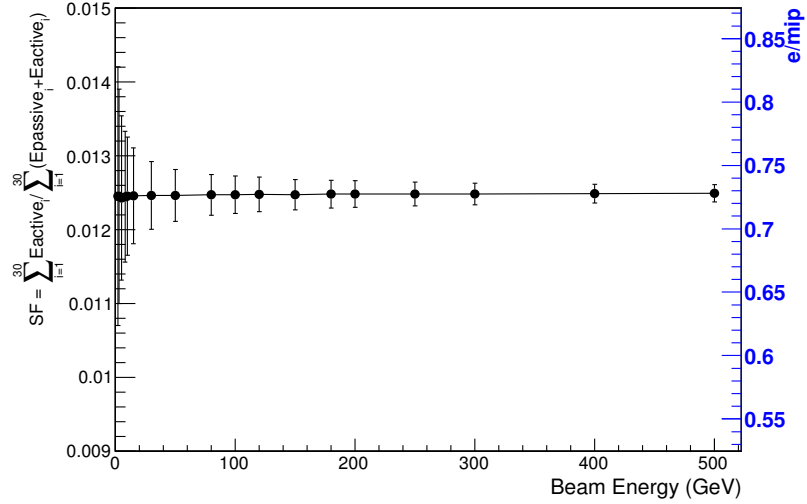}
}
\end{center}
\caption{
Total sampling fraction distribution as a function of incident
electron energy (upper graph) and mean value of the sampling fraction as a
function of energy (lower graph) for a homogeneous silicon sampling calorimeter.
In the lower graph, the means of the distributions are shown, while the error-bars correspond to the RMS of these distributions.
}
\label{sfhomog}
\end{figure}

\section{Results}

A silicon-pad sampling calorimeter is essentially a Minimum Ionizing Particle (MIP) counter. 
The simplest and most common method for energy reconstruction is the $dEdx$ method, in which the observed number of MIPs is weighted by a factor corresponding to the average energy lost per MIP in the passive layer. The flavour of $dEdx$ method employed here is as follows:
\begin{equation}
E_{rec} = \left( N_1\times\Delta E_{passive,1}^{MIP}
+   \Delta E_{silicon,1}\right) + \smashoperator[r]{\sum_{i=2}^{30}} \left(\frac{N_{i-1}+N_i}{2}\times\Delta E_{passive,i}^{MIP}
+   \Delta E_{silicon,i} \right),
\label{dedxformula}
\end{equation}
where $\Delta E_{passive}^{MIP}$ is the energy loss in a passive layer for a MIP, $\Delta E_{silicon}$ is the energy deposited in the silicon pad layer $i$, and $N_i$ the measured number of MIPs in the silicon pad layer $i$. This is an approximation 
not only because a particle at the minimum ionizing point will lose energy and eventually become non-MIP, but also because in the shower cascade process there are bremsstrahlung photons that may pair produce. Consequently, the method cannot provide the correct absolute energy scale. However, a good energy linearity is expected and in this case a simple rescaling to the absolute scale would suffice.\\

\begin{figure}[htb]
\begin{center}
\resizebox{0.8\textwidth}{!}{%
\includegraphics{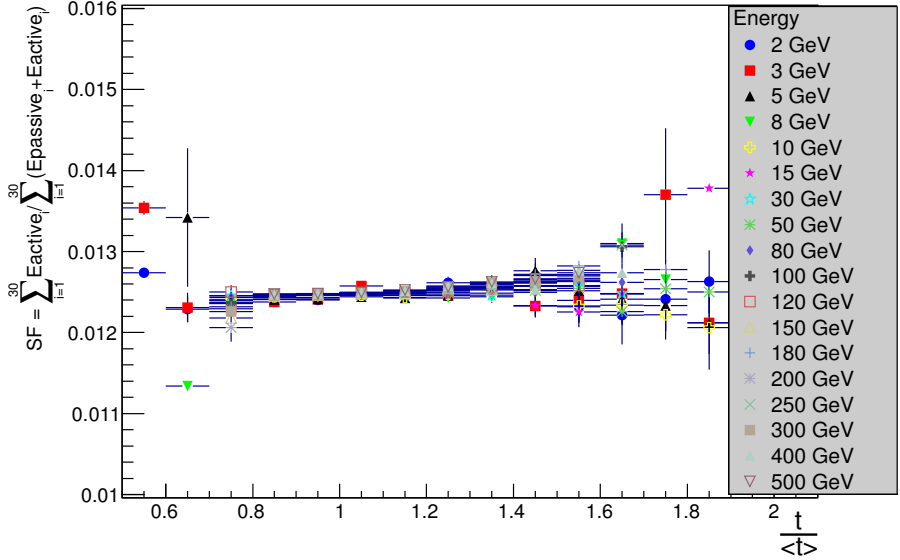}
}
\end{center}
\caption{
Dependence of mean value of sampling fraction on shower depth as a function of incident electron energy for a homogeneous silicon sampling calorimeter. 
}
\label{sfdepthhomog}
\end{figure}

\noindent
In this work, the total sampling fraction, $SF$, of a sampling calorimeter is defined as:
\begin{equation}
SF = \frac{ \smashoperator[r]{\sum_{i=1}^{30}} E_{active,i}
} 
{
\smashoperator[r]{\sum_{i=1}^{30}} E_{active,i} + 
\smashoperator[r]{\sum_{i=1}^{30}} E_{passive,i}
},
\end{equation}
where $E_{active}$ and $E_{passive}$ are the energies deposited in the active and passive layers, respectively. The typical value of $SF$ is of order of 1\%, which means that a correction of order 100 is needed to get to the true incident particle energy.
The total sampling fraction as a function of the incident electron energy for a silicon-pad sampling calorimeter with equal thickness passive layers (homogeneous calorimeter) is shown in Fig.~\ref{sfhomog}. 
The $e/mip$ ratio on the right vertical axis is estimated by dividing the $SF$ shown on the left vertical axis, by the sampling fraction for a MIP calculated using the $dEdx$ values in active and passive material:
\begin{equation}
e/mip=SF/\frac{ \smashoperator[r]{\sum_{i=1}^{30}} dEdx_{active,i}}{\smashoperator[r]{\sum_{i=1}^{30}} dEdx_{active,i} + \smashoperator[r]{\sum_{i=1}^{30}} dEdx_{passive,i}}.
\end{equation}
Although the $SF$ distribution gets narrower with energy, its mean value is constant, suggesting that a single calibration factor $SF^{-1}$ should be adequate for going from the measured energy to the reconstructed energy.  This is also supported by the dependence of $SF$ on shower depth normalized to the mean shower depth $t/<t>$, shown in Fig.~\ref{sfdepthhomog}. The shower depth $t$ is defined as:
\begin{equation} 
t = \frac{ \smashoperator[r]{\sum_{i=1}^{30}}\left( E_{i} \smashoperator[r]{\sum_{j=1}^{i}} X_{0,j}  \right) }{ \smashoperator[r]{\sum_{i=1}^{30}} E_{i}  }
\end{equation} 
and the mean $<t>$ is the mean value of the $t$ distribution for a particular energy. 
The indices $i$,$j$ refer to the calorimeter layers.
The $SF$ is almost independent of the shower depth,
meaning that the calorimeter response is the same, independent of longitudinal shower fluctuations.
Finally, the reconstructed energy using the corrected sampling fraction is: 
$E_{rec} = \smashoperator[r]{\sum_{i=1}^{30}} E_{active,i} \times SF^{-1}$.\\

\begin{figure}[htb]
\begin{center}
\resizebox{0.8\textwidth}{!}{%
\includegraphics{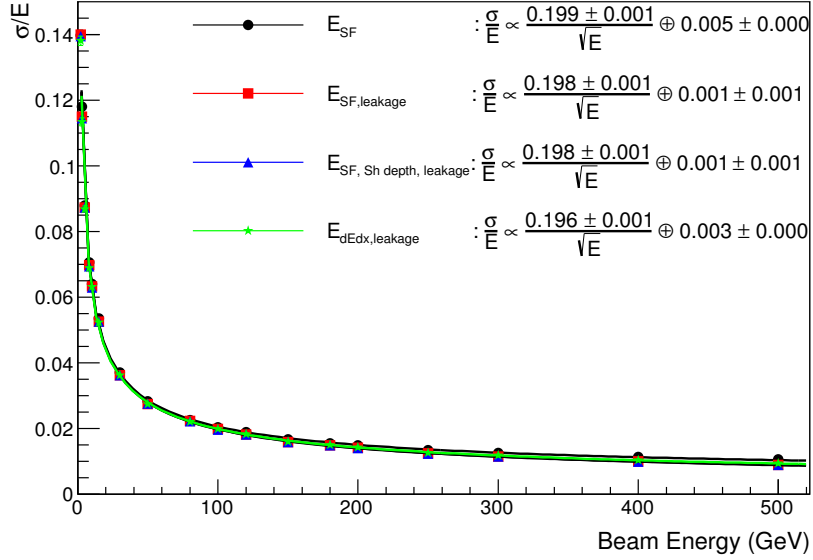}
}
\caption{
Energy resolution as a function of incident electron energy for a homogeneous silicon calorimeter.
When no correction for energy leakage from the back of the calorimeter is applied (black),
a non-zero constant term is introduced. If a correction is applied, then  
both $dEdx$ (green) and $SF$ (red and blue) methods have the same performance.
The $SF$ method corrected for shower-depth effects (blue) does not give any improvement 
in resolution, since the $SF$ is practically independent of the shower depth.
}
\label{erhomog}
\end{center}
\end{figure}

\begin{figure}[htb]
\begin{center}
\resizebox{0.8\textwidth}{!}{%
\includegraphics{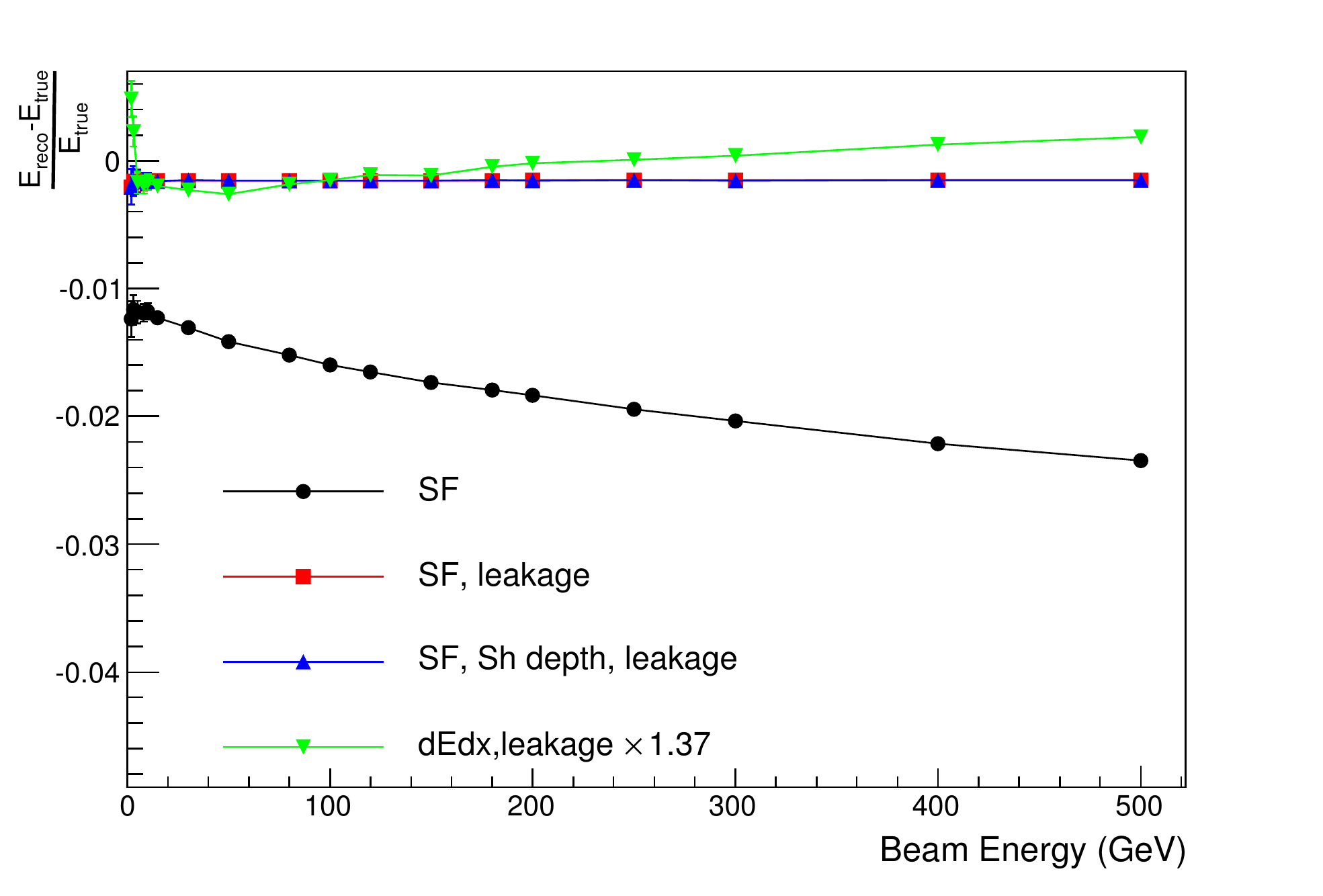}
}
\caption{
Energy linearity and scale of a homogeneous calorimeter as a function of incident electron energy using different calibration methods. The $dEdx$ method (green) undershoots the absolute energy scale by $\sim$37\% and has a mild non-linear behaviour. The $SF$ method (blue and red) is at the right energy scale. Not correcting for the leaking energy from the back of the calorimeter (black) induces non-linearity and scale effects.
}
\label{elhomog}
\end{center}
\end{figure}

\noindent 
In Figures~\ref{erhomog} and~\ref{elhomog}, the energy resolution and energy linearity performance of a homogeneous calorimeter are presented. 
The resolution is defined as 
$\sigma/E$, with $E$ and $\sigma$ the mean and sigma of a gaussian fit on the reconstructed energy.
The resolution versus energy graph is fitted with the function $a/\sqrt{E} + c$. In this analysis, no electronics noise is simulated. The final values for the stochastic term $a$ and the constant term $c$ can be seen in the graph. 
For the linearity plot in Figure~\ref{elhomog}, the absolute value of the mean and the error of the mean are obtained from the $\left( E_{reco}-E_{true}\right) /E_{true}$ distribution.
As shown, both the $dEdx$ and the $SF$ methods give similar results in terms of resolution, and this is to be expected, since in both cases a constant factor for all energies is applied. The $SF$ method gives the correct energy scale (by construction) while, it possesses a better linearity. In Fig.~\ref{elhomog}, the effect of not applying the energy leakage correction is also shown (black line). For the $dEdx$ method, the leakage correction is already applied.\\

\begin{figure}[htb]
\begin{center}
\resizebox{0.7\textwidth}{!}{
\includegraphics{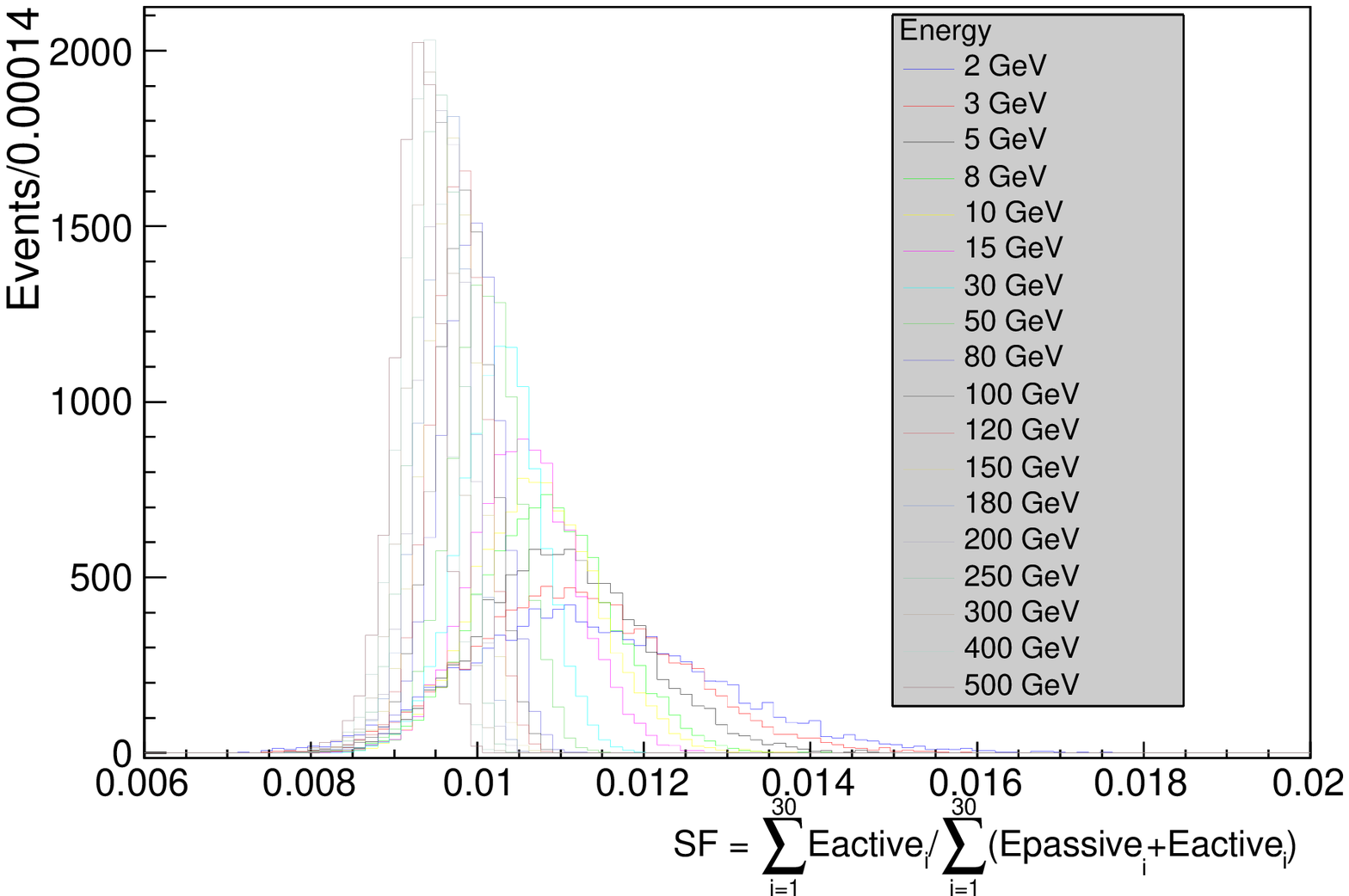}
}
\resizebox{0.72\textwidth}{!}{
\includegraphics{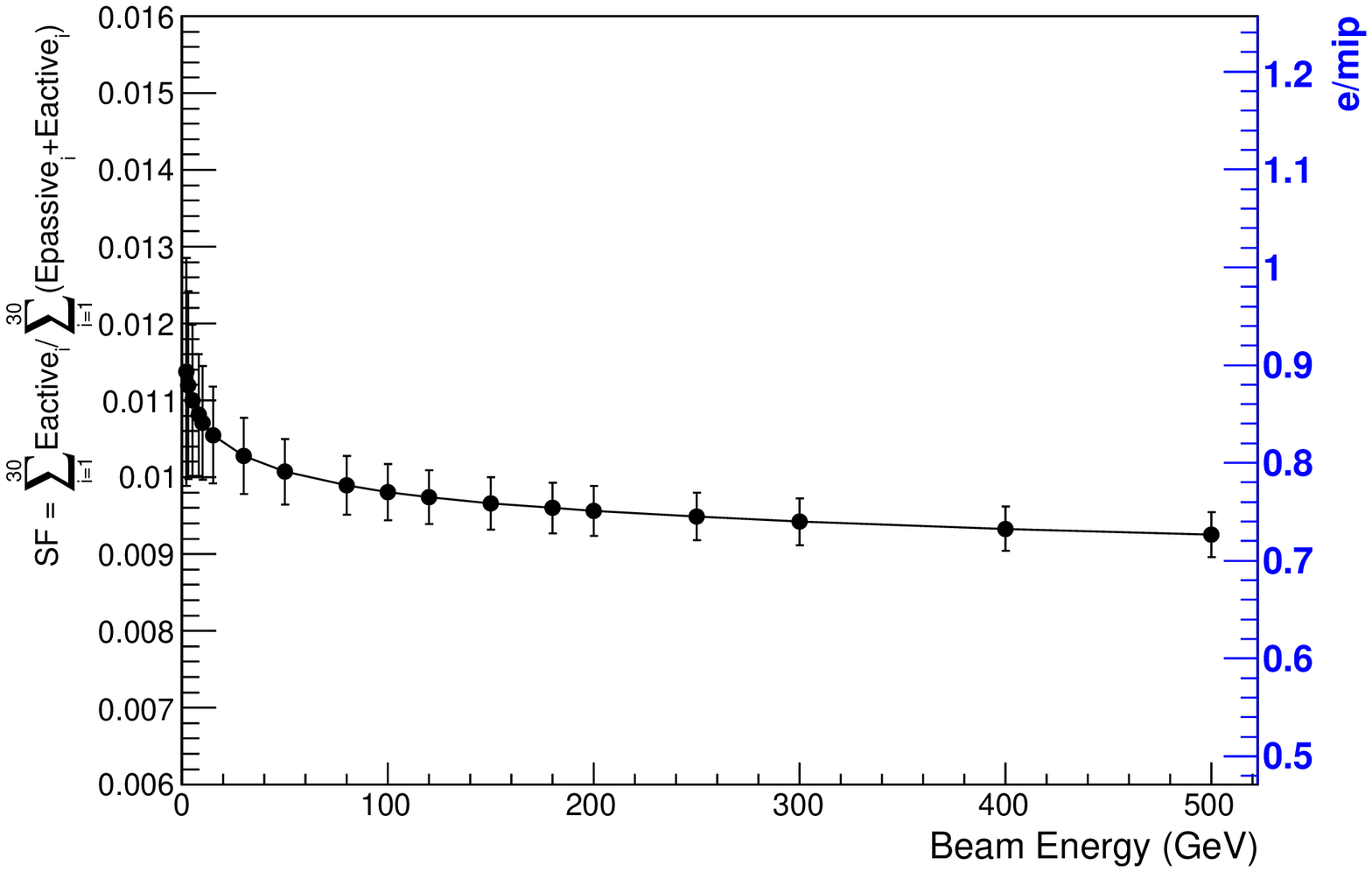}
}
\caption{
Total sampling fraction distribution as a function of incident
electron energy (upper graph) and mean value of the sampling fraction as a
function of energy (lower graph) for an inhomogeneous silicon sampling calorimeter.
The mean sampling fraction has an energy dependence.
}
\label{sfinhomog}
\end{center}
\end{figure}

\begin{figure}[htb]
\begin{center}
\resizebox{0.8\textwidth}{!}{%
\includegraphics{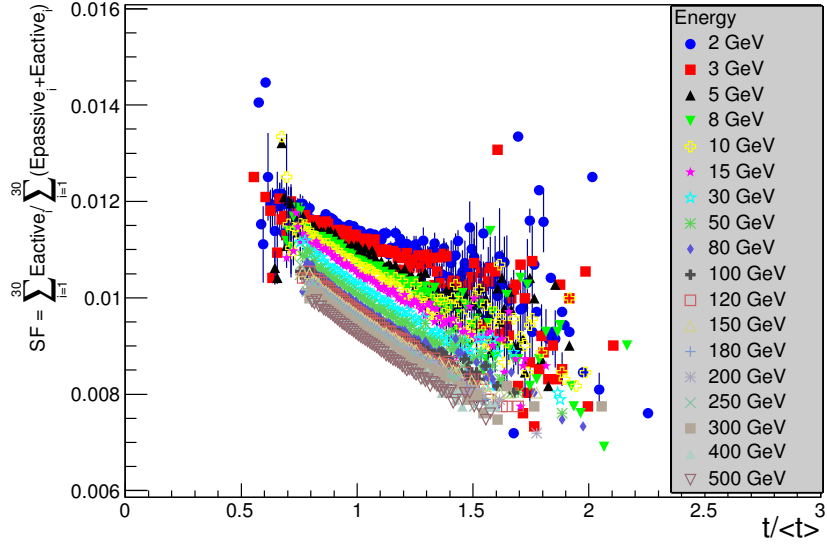}
}
\end{center}
\caption{
Dependence of the mean value of sampling fraction on shower depth as a function of incident electron energy for an inhomogeneous silicon sampling calorimeter. 
}
\label{sfdepthinhomog}
\end{figure}

\noindent
In contrast to the simple homogeneous design proposed for the next ILC and CEPC colliders, the first large scale detector using silicon pads as active material planned for CMS may be of an inhomogeneous or non-uniform design. Such a device could provide good performance for lower cost, as the back layers of the calorimeter have thicker passive layers and fewer channels per radiation length. Therefore, it is important to understand the level of compromise in performance, because such alternative lower-cost design paradigms can also be in the plans of future experiments. Inhomogeneous designs are typically designs of progressively increasing passive layer thickness, in particular towards the back of the calorimeter. In this case, later parts of the showers are sampled less, something that is not expected to give a significant performance loss in terms of energy reconstruction, provided that the core of the shower is still sampled at nominal, or even higher than nominal rate. Sampling less in the back of the calorimeter reduces the number of channels needed and this, besides reducing the cost, it may also lead to reduction of power consumption, and cooling requirements.
The main difference in response between such an inhomogeneous calorimeter and a homogeneous one, is the dependence of $SF$ on the shower depth.
The sampling fraction as a function of incident particle energy for an inhomogeneous calorimeter is shown in Fig.~\ref{sfinhomog}. The dependence of the $SF$ on shower depth for different energies is shown in Fig.~\ref{sfdepthinhomog}, where a clear reduction of the $SF$ is observed as a function of the normalized shower depth. This strong dependence on shower depth is due to the longitudinal increase of the passive layer thickness.
The main goal of this work is to examine the impact of such inhomogeneous geometry to calibration. In the case of the $SF$ method, it is now clear that $SF$ is a function of both energy and more importantly, shower depth. Attempting to calibrate using an average sampling fraction without correcting for the shower depth dependence will result to big loss of resolution. Fig.~\ref{sfdepthinhomog} suggests that this shower depth dependence is linear and universal, i.e. it has very little dependence on energy. Indeed, the negative slopes in the figure have very weak energy dependence for energies above 20~\GeV.  We can then use a single parameterization (i.e. a single slope) for the $SF$ as a function of shower depth, independent of the incident particle energy to proceed with the calibration. In such fine longitudinally segmented calorimeters the shower depth is an accurately measured quantity, so in practice this is a simple correction to apply in the first stage of a calibration programme.
The corrected sampling fraction is obtained through a linear fit of the $<SF>=f\left( t/<t>\right)$ relation, Fig.~\ref{sfdepthinhomog}, as follows:
\begin{equation}
SF_{corr}=\lambda \times \left( \frac{t}{<t>}\bigg\rvert_{l} - \frac{t}{<t>}\bigg\rvert_{bin} \right) + <SF>\bigg\rvert_{bin},
\label{SFcorr}
\end{equation}
where $\frac{t}{<t>}\rvert_{l}$ is the measured value of the normalized shower depth, $\frac{t}{<t>}\rvert_{bin}$ is the mean value of the shower depth for the measured event energy,  $<SF>\rvert_{bin}$ is the mean sampling fraction for that energy, and $\lambda$ is the universal slope from the fit.\\

\begin{figure}[htb]
\begin{center}
\resizebox{0.8\textwidth}{!}{%
\includegraphics{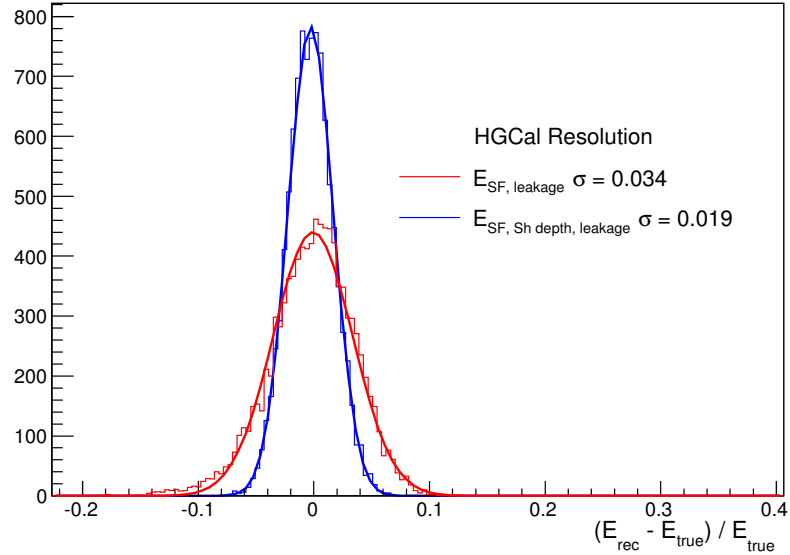}
}
\caption{
Energy resolution before and after correcting for the $SF$ dependence on the shower depth in an inhomogeneous silicon sampling calorimeter for incident electrons of 100~\GeV ~energy. 
}
\label{impactResol}
\end{center}
\end{figure}

\noindent
The impact of the event-by-event shower shape correction on energy resolution is shown in Fig.~\ref{impactResol}. As expected, in the case of an inhomogeneous calorimeter the use of an average sampling fraction does not give the optimum resolution due to the strong $SF$ dependence on shower depth. In Fig.~\ref{impactResol}, we correct the $SF$ according to the measured shower depth and obtain an event-by-event corrected reconstructed energy according to Eq.~\ref{SFcorr}. This correction leads to a resolution close to that of the homogeneous calorimeter.\\

\begin{figure}[htb]
\begin{center}
\resizebox{0.8\textwidth}{!}{%
\includegraphics{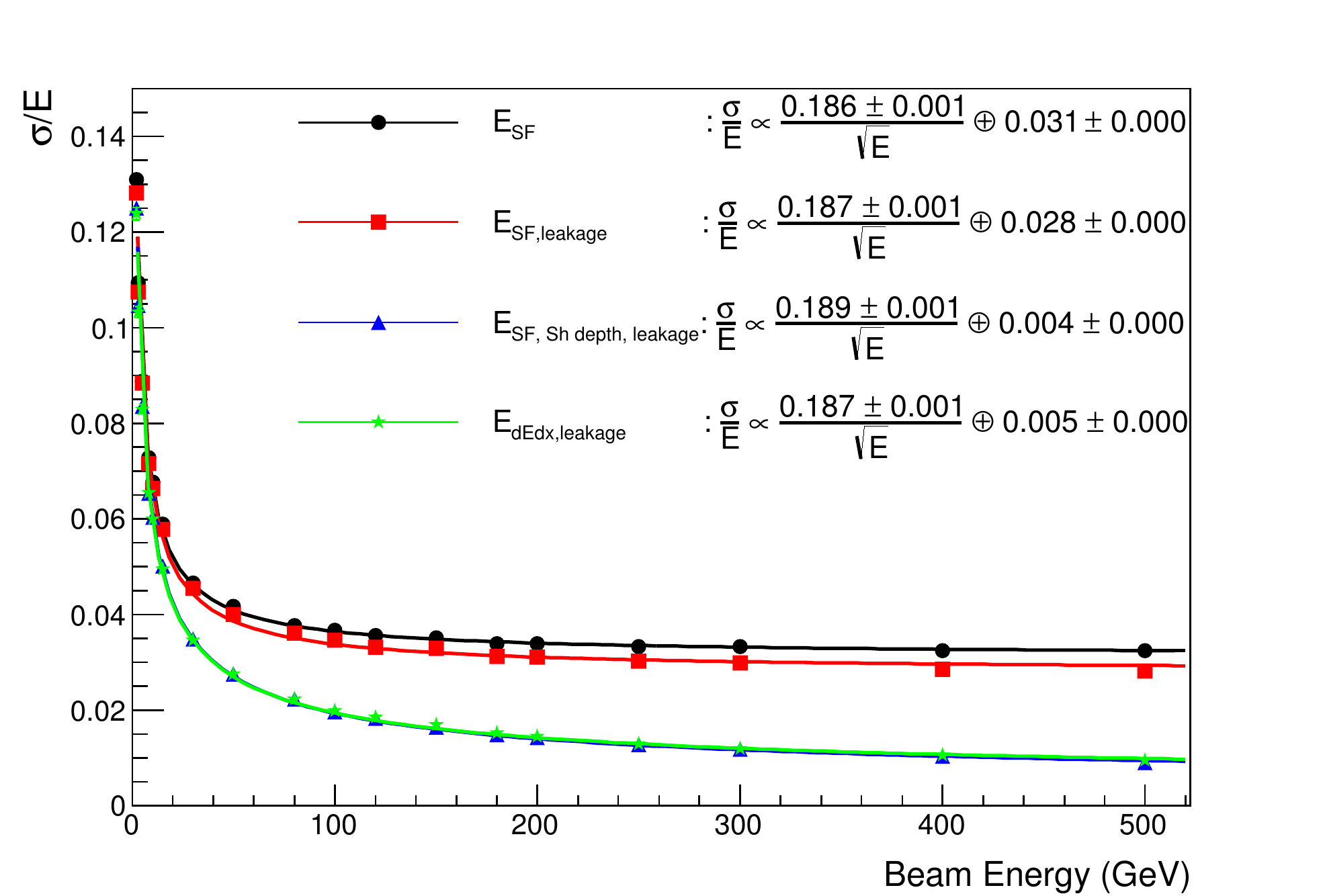}
}
\caption{
Energy resolution of an inhomogeneous silicon calorimeter as a function of incident electron energy using different calibration methods.
The $SF$ method with a universal linear correction and a per-event correction using the measured shower depth (blue) gives a similar stochastic term as the standard $dEdx$ method (green). However, the $SF$ method gives a significantly reduced constant term. It should be noted that at low electron energies below 50~\GeV ~there are departures from universality, so the shower-depth correction could be further improved.
}
\label{erinhomog}
\end{center}
\end{figure}

\begin{figure}[htb]
\begin{center}
\resizebox{0.8\textwidth}{!}{%
\includegraphics{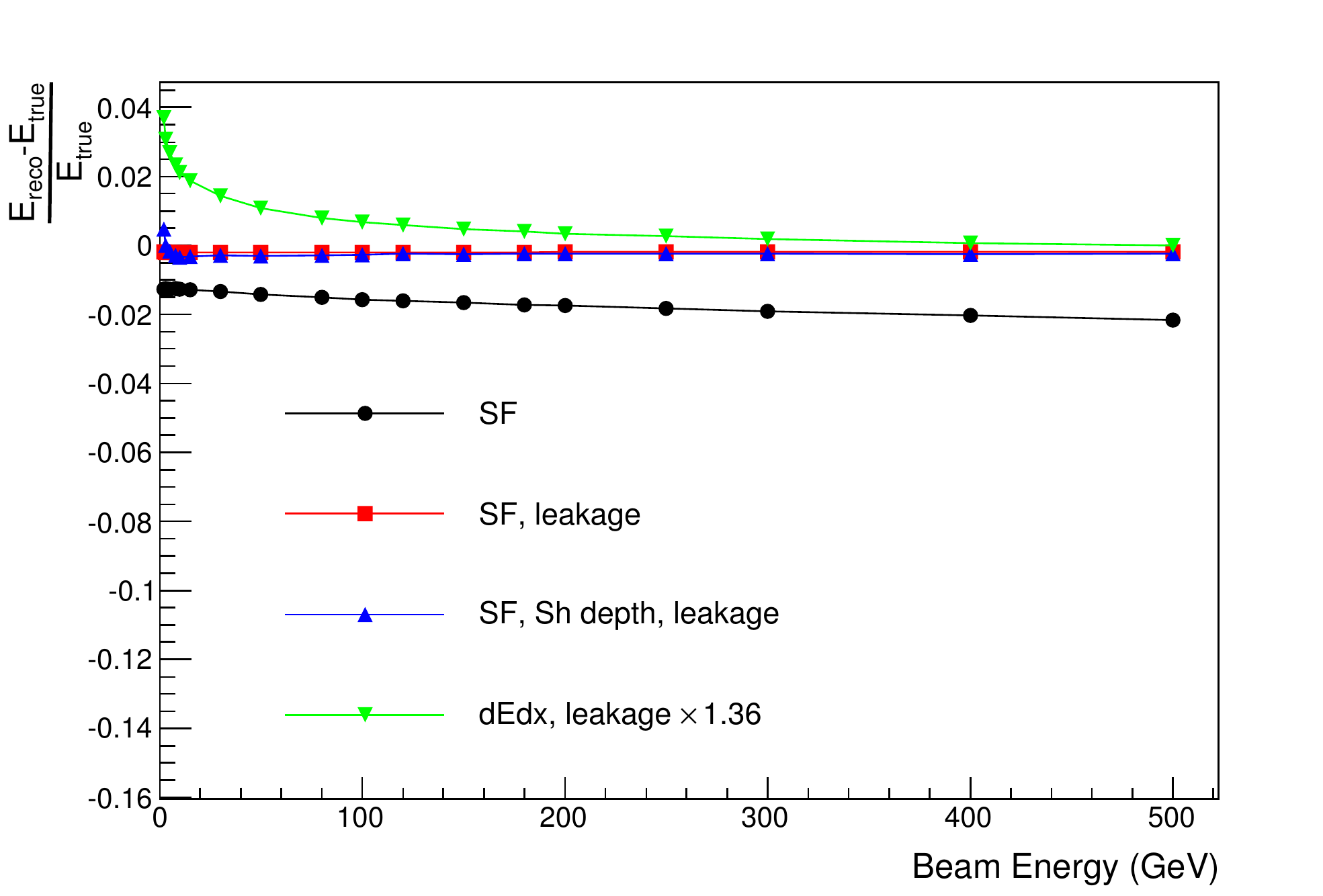}
}
\caption{
Energy linearity of an inhomogeneous silicon calorimeter as a function of incident electron energy using different calibration methods. The $SF$ method (blue and red) is linear, while at the same time provides the correct energy scale. The $dEdx$ method (green) undershoots the absolute energy scale by $\sim 36$\% and shows departure from linearity, in particular for energies below 100~\GeV. The $SF$ calibration without leakage correction is also shown in absolute value of the linearity variable (black).
}
\label{elinhomog}
\end{center}
\end{figure}

\noindent
The energy resolution of an inhomogeneous silicon calorimeter as a function of the incident electron energy using different calibration methods is shown in Fig.~\ref{erinhomog}.
The sampling-fraction method with a universal linear per-event correction using the measured shower depth, gives a similar stochastic term as the $dEdx$ method. However, the $SF$ method gives a constant term reduced by 20\%. It should be noted that at low electron energies below 50~\GeV ~there are departures from universality, so the correction could be further improved.
Finally, the energy linearity of an inhomogeneous calorimeter as a function of incident electron energy using different calibration methods is shown in Fig.~\ref{elinhomog}. The $SF$ method is linear while at the same time provides the correct energy scale. The $dEdx$ method undershoots the absolute energy scale by about 36\% and shows departure from linearity, in particular for energies below 100~\GeV.
The  reconstructed energy in the $dEdx$ method is calibrated to the absolute
energy scale using a reference energy of 500~\GeV ~in Fig.~\ref{elinhomog}.
Starting instead with a $SF$ approach that provides a better starting point on the absolute energy scale and linearity for a slightly better resolution, would be the preferred choice for an initial calibration of the detector.

\section{Conclusions}
In this work, the energy resolution, linearity and scale of a realistic silicon-based sampling calorimeter was studied. It was shown that instead of the default $dEdx$ method, a method based on the sampling fraction gives superior performance, in particular in the case of an inhomogeneous calorimeter for which the layer thickness varies longitudinally. Such designs could be cost-effective and have already been proposed for the CMS experiment for the LHC phase-2 programme. Unlike the $dEdx$ method, the sampling fraction method leads to a good energy linearity for the full energy range and a 20\% reduced residual constant term of 0.4\%. Inhomogeneous sampling calorimeters can also be employed in large future detectors for ILC and CEPC, as well as in particle astrophysics satellite mission experiments to reduce costs and services. The work presented here will be useful in the choice of detector designs with optimal performance and cost.\\

\noindent
The next essential step would be to demonstrate the studies presented in this work in silicon calorimeter test-beam campaigns with a complete EM prototype.

\acknowledgments
The authors would like to thank Valeri Andreev (CERN) for suggesting the use of the $dEdx$ formula 
of Eq.~\ref{dedxformula}.


\end{document}